\documentstyle[aps,amsmath,amssymb,10pt,psfig]{revtex}

\begin{document}

\draft

\newcommand{\nc}{\newcommand}
\nc{\ww}{\omega}

\title{ HeXLN: A 2-Dimensional nonlinear photonic crystal} 
\author{ N. G. R. Broderick, G.~W.~Ross, H.~L.~Offerhaus, D.~J.~Richardson 
		and D.~C.~Hanna} 
\address{Optoelectronics Research Centre, University of Southampton,
      Southampton, SO17 1BJ, UK. \\
      Phone: +44  (0)1703 593144, Fax: +44 (0)1703 593142,
      email:  ngb@orc.soton.ac.uk} 

\date{\today}

\maketitle

\begin{abstract}
We report on the fabrication of what we believe is the first example of a two 
dimensional nonlinear
periodic crystal\cite{berger}, where the refractive index is constant 
but in which the 2nd order nonlinear susceptibility is spatially periodic. 
Such a crystal allows for efficient quasi-phase matched 2nd harmonic 
generation using multiple reciprocal lattice vectors of the crystal 
lattice. External 2nd harmonic conversion efficiencies $> 60\%$ were 
measured with picosecond pulses. The 2nd harmonic light can be 
simultaneously phase matched by multiple reciprocal lattice 
vectors, resulting in the generation of multiple coherent beams. The 
fabrication technique is extremely versatile and allows for the 
fabrication of a broad range of 2-D crystals including quasi-crystals. 
\end{abstract}


The interaction of light with periodic media is an area of intense interest
both theoretically and experimentally. A central theme of this work is the
idea of a linear photonic crystal\cite{photonic} in which the linear 
susceptibility  is spatially periodic. Photonic crystals can have a 
complete photonic bandgap over some frequency range and this bandgap can be 
exploited 
for a wide variety of processes such as zero threshold lasers, inhibited 
spontaneous emission, or novel waveguiding schemes such as photonic bandgap 
fibres\cite{bath1}. In one dimension photonic crystals, or Bragg gratings, 
have been well studied for many years. In three
dimensions a complete photonic bandgap at long wavelengths has already been
demonstrated and work on extending this to the visible region is rapidly 
progressing\cite{photonic}. 

Recently V.~Berger proposed extending the idea of photonic crystals to  
include nonlinear photonic crystals\cite{berger}.
In a nonlinear photonic crystal (NPC) there is a periodic spatial variation
of a nonlinear susceptibility tensor
while the refractive index is constant. This is in 
contrast with other work done on nonlinear interactions in photonic 
crystals\cite{sajeev,mike2nd} where the nonlinearity is 
assumed constant throughout 
the material and the photonic properties derive from the variation of the linear
susceptibility. The simplest type of NPCs are the 1-D quasi-phase-matched  
 materials, first proposed by Armstrong {et al.}\cite{qpm1} in which the
second order susceptibility undergoes a periodic change of sign. This type of
1-D structure has attracted much interest since the successful development of 
periodically-poled lithium niobate based devices. Generalisation to two and 
three dimensions in analogy with  linear photonic crystals, was recently 
proposed by Berger and here we report its experimental realisation 
as a 2-D periodic structure with hexagonal symmetry in lithium niobate (HeXLN).

First we briefly summarise the well known 1-D quasi-phase matching (QPM)
 concept before treating the
2-D case. To this end consider the case of 2nd harmonic generation in
a $\chi^{(2)}$ material where light at a frequency $\ww$ is converted to a
signal at $2\ww$. In general the 
refractive index at $\ww$ and $2\ww$ are different and hence after a length
$L_c$ (the coherence length) the fundamental 
and the generated 2nd harmonic will be $\pi$ out of phase. Then in the next 
coherence length  all of the 2nd harmonic is back-converted to the 
fundamental - resulting 
in poor overall conversion efficiency. The idea of quasi-phase matching is to
change the sign of the nonlinearity periodically with a period of $L_c$, thus
periodically reversing the phase of the generated 2nd harmonic. This
ensures that the 2nd harmonic continues to add up in phase along the entire
length of the crystal, resulting in a large overall conversion efficiency. 

An alternative way to understand the physics of quasi-phase matching is 
through conservation of momentum. 2nd harmonic generation is a three photon
process in which two photons with momentum $\hbar k^\ww$ are converted in a 
photon of momentum $\hbar k^{2\ww}$ and if $k^{2 \ww} = 2 k^\ww$ (ideal 
phase matching) then the momentum is conserved and the interaction is efficient. 
However in general due to dispersion ideal phase matching is not possible and
different techniques must be used to insure conservation of momentum.  In the 
quasi-phase matched
case conservation of momentum becomes $k^{2 \ww} = 2 k^\ww +G,$ where $G$ is
the crystal momentum corresponding to one of the reciprocal lattice vectors
(RLV)\cite{kittel} of the macroscopic periodic structure of the NPC.
 Clearly this technique allows one to phase-match any desired 
nonlinear interaction, assuming that one can fabricate an appropriate NPC. 
In 1-D quasi-phase matching can  occur in either the co- or 
counter-propagating direction. For a strictly 
periodic lattice quasi-phase matching can only occur over limited 
wavelength ranges since the RLVs are discrete and periodically spaced in 
momentum space. In order to obtain broader bandwidths one approach is to
use aperiodic structures which have densely spaced RLVs. An alternative
approach which is taken here is to move to a two dimensional 
NPC which brings added functionality compared to a 1-D crystal.

Clearly in a 2-D NPC the possibility of non-collinear phase matching exists 
due to the structure of the reciprocal lattice.  Once again  we restrict
ourselves to the case of 2nd harmonic generation and linearly polarised
light such that we can use the scalar wave equation. Then  
making the usual slowly varying envelope 
approximation and assuming a plane wave fundamental incident upon the 
crystal, the evolution equation for the 2nd 
harmonic in the undepleted pump regime can be written as\cite{berger}: 
\begin{equation}
\label{wave1}
{\bf k}^{2\ww}\cdot \nabla E^{2 \ww}({\bf r}) = -2 i \frac{\ww^2}{c^2}
\chi^{(2)}({\bf r}) (E^\ww)^2 
e^{\left(i ({\bf k}^{2 \ww} - 2 {\bf k}^\ww)\cdot {\bf r}\right)}.
\end{equation}
Since $\chi^{(2)}$ is periodic we can write it as a Fourier series using the 
RLVs ${\bf G}_{n,m}$ 
\begin{equation}
\label{kaps}
	\chi^{(2)}({\bf r}) = \sum_{n,m} \kappa_{n,m} e^{i {\bf G}_{n,m}
\cdot {\bf r} }, \quad n,m \in {\Bbb Z}.
\end{equation}
 The phase matching condition, 
\begin{equation}
\label{qpm}
{\bf k}^{2\ww} -2{\bf k}^\ww  -{\bf G}_{n,m}=0, 
\end{equation}
arises from requiring that the exponent in Eq.~\eqref{wave1} be set equal to
zero ensuring growth of the 2nd harmonic along the entire length of the 
crystal.  Eq.~\eqref{qpm} is a 
statement of conservation of momentum as discussed earlier. For each RLV 
${\bf G}_{n,m}$ and a prescribed ${\bf k}^\ww$ 
 there is at most a unique angle of propagation for the 2nd 
harmonic such that Eq.~\eqref{qpm} is satisfied. The coupling strength
 of a phase matching process using ${\bf G}_{n,m}$ is proportional to 
$\kappa_{n,m}$. If a particular Fourier coefficient is zero then 
no 2nd harmonic generation will be observed in the corresponding direction.  

In order to demonstrate the idea of a 2-D NPC we poled a wafer of 
lithium niobate with a hexagonal pattern.  
Fig.~\ref{hexln} shows an expanded view of the resulting structure, which was
revealed by lightly etching the sample in acid. Each hexagon is a region of 
domain inverted material - the total inverted area comprises $\sim 30\%$ of the 
overall sample area. The fabrication procedure was as follows.
A thin layer of photoresist was first deposited onto the -z face of a
0.3mm thick, z-cut wafer, of LiNb${\rm O}_3$, and then photolithographically
patterned with the hexagonal array. The x-y orientation of the
hexagonal structure was carefully aligned to coincide with the
crystal's natural preferred domain wall orientation : LiNb${\rm O}_3$ itself
has  triagonal atomic symmetry (crystal class 3m) 
and shows a tendency for domain walls
to form parallel to the y-axis and at $\pm60^\circ$ as seen in
Fig.~\ref{hexln}. Poling was accomplished
by applying an electric field via liquid electrodes on the +/-z faces at
room temperature\cite{yamada}.  Our HeXLN crystal has a period of 
$18.05\, \mu$m: suitable for
non-collinear frequency doubling of 1536nm at 150$^\circ$C (an elevated 
temperature was chosen to eliminate photorefractive effects). 
The hexagonal pattern 
was found to be uniform across the sample dimensions of 14 $\times$ 7mm (x-y)
and was faithfully reproduced on the +z face.  Lastly we polished the $\pm x$
-faces of the HeXLN crystal
allowing a propagation length of 14mm through the crystal in the
$\Gamma{\rm K}$ direction (see Fig.~\ref{hexln}).

In Fig.~\ref{lattice} we show the reciprocal lattice (RL) for our HeXLN crystal.
In contrast with the 1-D case there are RLVs at numerous angles, each of 
which allows phase matching in a different direction 
(given by Eq.~\ref{qpm}).
Note that for a real space lattice period of $d$ the RL has a period of
$4\pi/(\sqrt{3} d)$ as compared with $2 \pi/d$ for a 1-D crystal\cite{kittel2}
allowing us to compensate for a greater phase mismatch in a 2-D geometry
than in a 1-D geometry with the same spatial period. From Eq.~\eqref{qpm} 
and using simple trigonometry it is possible to show that\cite{berger}
\begin{equation}
\label{angles}
\frac{\lambda^{2 \ww}}{n^{2\ww}} = \frac{2 \pi}{|{\bf G}|} 
\sqrt{\left(1-\frac{n^\ww}{n^{2 \ww}}
\right)^2 + 4 \frac{n^\ww}{n^{2 \ww}}\sin^2\theta }
\end{equation}
where $\lambda^{2 \ww}$ is the vacuum wavelength of the second harmonic 
and  $2 \theta$ is the walk off angle between the fundamental and 
2nd harmonic wavevectors. 

To investigate the properties of the HeXLN crystal we proceeded as follows. 
The HeXLN crystal was placed in an oven and mounted on a rotation stage 
which could be rotated by $\pm 15^\circ$ around the z-axis 
while still allowing light to enter through the $+x$ face of the crystal. 
The fundamental consisted of 4ps, 300kW  pulses obtained from a high power 
all-fibre chirped pulse amplification system (CPA)\cite{neilcpa} operating at  a
pulse repetition rate of 
20kHz. The output from the CPA system was focussed into the HeXLN crystal 
using a 10cm focal length lens giving a focal spot diameter of 
$150\mu{\rm m}$ and a corresponding peak intensity of 
$\sim 1.8$GW/${\rm cm}^2$. 
The initial experiments were done at zero angle of incidence corresponding to
propagation in the $\Gamma {\rm K}$ direction. At low input intensities 
($\sim 0.2$GW/${\rm cm}^2$) the output was as shown in 
Fig.~\ref{lattice}(b) and consisted of multiple output
beams of different colours emerging from the crystal at different angles. 
In particular two 2nd harmonic beams emerged from the crystal at symmetrical
angles of $\pm (1.1 \pm 0.1)^\circ$ from the remaining undeflected fundamental. 
At slightly wider 
angles were two green beams (third harmonic of the pump) and at even wider 
angles were two blue beams (the fourth harmonic, not shown here). There was 
also a third green beam copropagating with the fundamental. The output was 
symmetrical since
the input direction corresponded to a symmetry axis of the NPC. As 
the input power increased the 2nd harmonic spots remained in the same 
positions while the green light appeared to be emitted over an almost 
continuous range of angles rather than the discrete angles observed at low 
powers.  The two 2nd harmonic beams can be understood by referring to the 
reciprocal lattice  of our structure (Fig.~\ref{lattice}). From 
Fig.~\ref{lattice} it can be seen that for propagation in the $\Gamma K$ 
direction the closest RLVs are in the $\Gamma M$ directions 
 and it is these RLVs that account for the 2nd harmonic light\cite{berger}. 

After filtering out the other wavelengths the 2nd harmonic (from both beams) 
was directed onto a power meter and the efficiency and temperature tuning 
characteristics for zero input angle were 
measured. These results are shown in Fig.~\ref{eff} and Fig.~\ref{temp}. 
Note that the maximum  
external conversion efficiency is greater than $60\%$ and this is 
constant over a wide range of input powers. Taking into account the 
Fresnel reflections 
from the front and rear faces of the crystal this implies a maximum internal
conversion efficiency of $82\%$ -- $\sim 41\%$ in each beam. As the 2nd 
harmonic power increases the amount of back conversion increases which we
believe is the main reason for the observed limiting of the conversion 
efficiency at high powers. 

Evidence of the strong back conversion can be seen in Fig.~\ref{spectra}
which shows the spectrum of the remaining fundamental 
for both vertically (dashed) i.e. in the z-direction and horizontally 
(solid line) polarised input light.
As the phase matching only works for vertically polarised light the
horizontally polarised spectrum is identical to
that of the input beam and when compared with the other trace (dashed line)
shows the effect of pump depletion and back-conversion. Note that for
vertically polarised light the amount of back-converted light is significant
compared to the residual pump which is as expected given the large
conversion efficiency.  Fig.~\ref{spectra} shows $\sim 8$dB ($85\%$) of pump
depletion which agrees well with the measured value for the internal efficiency
calculated using the average power.

In the 1-D case, for an undepleted pump, the temperature tuning curve of a 
14mm long length of periodically
poled material is expected to have a $sinc^2(T)$ shape and to be quite 
narrow -- $4.7^\circ$C  for a 1-D PPLN crystal with the same length
and period as the HeXLN crystal used here. However, as can be seen from 
Fig.~\ref{temp}, the temperature tuning curve (obtained in a similar 
manner to the power characteristic) is much broader with a FWHM of 
$\sim 25^\circ$C, and it exhibits considerable structure. The input
power was 300kW. 
We believe that the increased temperature bandwidth may be due to the multiple 
reciprocal lattice vectors that are available for quasi-phase matching with 
each RLV producing a beam in a slightly different direction. Thus the angle
of emission of the 2nd harmonic should vary slightly with temperature if this 
is the case. Due to the limitations of the oven we 
were not able to raise the temperature above $205^\circ$C and hence 
could not completely measure the high temperature 
tail of the temperature tuning curve. 
 Note that temperature tuning is equivalent to 
wavelength tuning of the pump pulse and hence it should be possible to 
obtain efficient phase-matching over a wide wavelength range at a fixed 
temperature.  

After the properties of the HeXLN crystal at normal incidence we next measured
the angular dependance of the 2nd harmonic beams.
As the crystal was rotated phase-matching via different RLVs could be
observed. For a particular input angle (which determined the angle between
the fundamental and the RLVs) quasi-phase matched 2nd harmonic generation
 occurred, via a 
RLV, and produced a 2nd harmonic beam in a direction given by
Eq.~\eqref{angles}. These results are shown in Fig.~\ref{figangles} where
the solid circles indicate the measured angles of emission for 2nd harmonic while
the open squares are the predicted values. In the figure zero degrees 
corresponds to propagation in the $\Gamma K$ direction.
 Also indicated on the figure are the RLVs used for phase-matching, where
$\left[n,m\right]$ refers to the RLV ${\bf G}_{n,m}$. Note that
there is good overall agreement between the theoretical and experimental
results even for higher order Fourier coefficients which indicates the
high quality of the crystal. The inversion symmetry of Fig.~\ref{figangles}
results from the hexagonal symmetry of the crystal. To further
highlight this symmetry we have labeled the negative
output angles with the corresponding positive RLVs. The only obvious
discrepancy comes from the $\left[1, 1\right]$ RLVs where two closely
separated spots are observed rather than a single one. This may be
 due to a small amount of linear diffraction from the
periodic array. At the domain boundaries of the HeXLN crystal there are likely
to be small stress-induced refractive index changes giving a
periodic variation in the refractive index. If this indeed proves to be the
 case then it should be
possible to eliminate this by annealing the crystal at high temperatures.

For applications where collinear propagation of the fundamental and 2nd
harmonic is desirable propagation along the $\Gamma M$
axis of the HeXLN crystal could be used (since the smallest RLV is in that
direction). For the parameters of our crystal
collinear 2nd harmonic generation of 
$1.446\,\mu{\rm m}$ in the $\Gamma M$ direction is expected. 

Visually the output of the HeXLN crystal is quite striking with 
different colours (red, green and blue) being emitted in different directions
(see Fig.~\ref{output}).
For a range of input angles and low powers distinct green and 
red spots can been seen each emitted in a different direction, often with
the green light emitted at a wider angle than the 2nd harmonic. 
The presence of the green light implies sum frequency 
generation between the fundamental and the 2nd harmonic. For this to occur 
efficiently it must also be quasi-phase-matched using a 
RLV of the lattice. In certain regimes (of angle and temperature) 
simultaneous quasi-phase-matching of both 2nd harmonic generation and 
sum frequency mixing occurs with as much as 20\% of the 2nd harmonic, 
converted to the green (in multiple beams). As mentioned 
earlier at higher powers the green light appears to be emitted over a 
continuous range of angles. We believe that this might be due to an effect
similar to that  observed in fibres where phase-matching becomes less 
critical at high intensities\cite{trillo}. If this were the case then the 
green light would have a broader spectrum in the non-phase-matched case than
for the quasi-phase-matched case but we have not yet been able to verify 
this.  Lastly we believe that the 4th harmonic results from quasi-phase 
matching of two 2nd harmonic photons by a higher order RLV since it is 
observed at quite wide angles. 

It should be noted that although lithium niobate preferentially forms domains
walls along the $y$ axis and at $\pm 60^\circ$ we are not limited to 
hexagonal lattices. In fact essentially any two dimensional lattice can be 
fabricated,  however the patterned region of the unit cell  will always
consist of either a hexagon or a triangle. The shape of the poled region 
will determine the strength of each of the Fourier coefficients for the RLVs 
while the lattice structure will determine their position. 
One can envisage creating more complicated structures such as a 2-D 
quasi-crystal in which a small poled hexagon is situated at every vertex. 
Such a 2-D quasi-crystal could give improved performance for simultaneously 
phase matching multiple nonlinear processes, as demonstrated recently with a
1-D poled quasi-crystal\cite{fib}. Alternatively a HeXLN crystal could be used 
as an efficient monolithic optical parametric oscillator\cite{berger}.  
Lastly we note that NPCs are a specific example of  more general nonlinear
holographs which would convert a beam profile at one wavelength to an arbitrary
profile at a second profile\cite{berger2}. For example Imeshevx {\it et al.}
converted a gaussian profile beam at the fundamental to a square top 2nd 
harmonic using tranversely patterned periodically poled lithium 
niobate\cite{fejer2d}.

In conclusion we have fabricated what we believe to be the first example of 
a two dimensional nonlinear photonic crystal in Lithium Niobate. Due to the
periodic structure of the crystal, quasi-phase matching is obtained 
for multiple directions of propagation with internal conversion efficiencies 
of $> 80\%$. Such HeXLN crystals could find many 
applications in optics where simultaneous conversion of multiple wavelengths 
is required.

\begin{figure}[h]
\caption{\label{hexln} Picture of the HeXLN crystal and the first Brillouin
zone.  The period of the crystal is 18.05$\mu$m and is uniform over the whole
sample. In our experiments propagation was in the $\Gamma K$ direction.
}
\end{figure}

\begin{figure}[h]
\caption{\label{lattice} Reciprocal Lattice for the hexagonal lattice shown
in Fig.~\ref{hexln}. The general reciprocal lattice vector ${\bf G}_{n,m} =
n {\bf e}_1 + m {\bf e}_2$ where ${\bf e}_{1,2}$ are the basis vectors for
the reciprocal lattice.  Also indicated is the first Brillouin zone showing the
main symmetry directions. In addition two examples of non-collinear QPM are
shown using the $\left[1,0\right]$ and the $\left[1,1\right]$ RLVs. 
On the right is a picture of the low power output of the crystal. Note that
there are two 2nd harmonic spots and three 3rd harmonic spots. 
}
\end{figure}

\begin{figure}[h]
\caption{\label{eff}  2nd harmonic efficiency of the HeXLN
crystal against input peak power. Note that the maximum efficiency is 
$> 60\%$ and is limited principally by parametric back conversion. }
\end{figure}

\begin{figure}[h]
\caption{\label{spectra}  Output spectra at 1533nm for both horizontally (solid
line) and vertically (dashed line) polarised light.  Note the large amount of
pump depletion which can clearly be seen along with the back-conversion. The
incident peak power was 300kW.
}
\end{figure}

\begin{figure}[h]
\caption{\label{temp}  Temperature tuning of the HeXLN
crystal taken at an incident peak power of 300kW. The temperature tuning
curve is much broader than a comparable 1-D PPLN crystal and posses multiple
features has to the large number of reciprocal lattice vectors available.
}
\end{figure}

\begin{figure}[h]
\caption{\label{figangles}
Graph of the experimental (circles) and theoretical (squares) output angles
for the 2nd harmonic as an function of the external input angle, where
$0^\circ$ indicates propagation in the $\Gamma {\rm K}$ direction. The maximum
internal
angle between the fundamental and 2nd harmonic was $\sim 8^\circ$ (the
refractive index of lithium niobate  is $\sim$ 2.2).
}
\end{figure}

\begin{figure}[h]
\caption{\label{output} Output of the HeXLN crystal at high powers and a 
variety of input angles.}
\end{figure}

\newpage

\begin{figure}[h]
\centerline{
\psfig{file=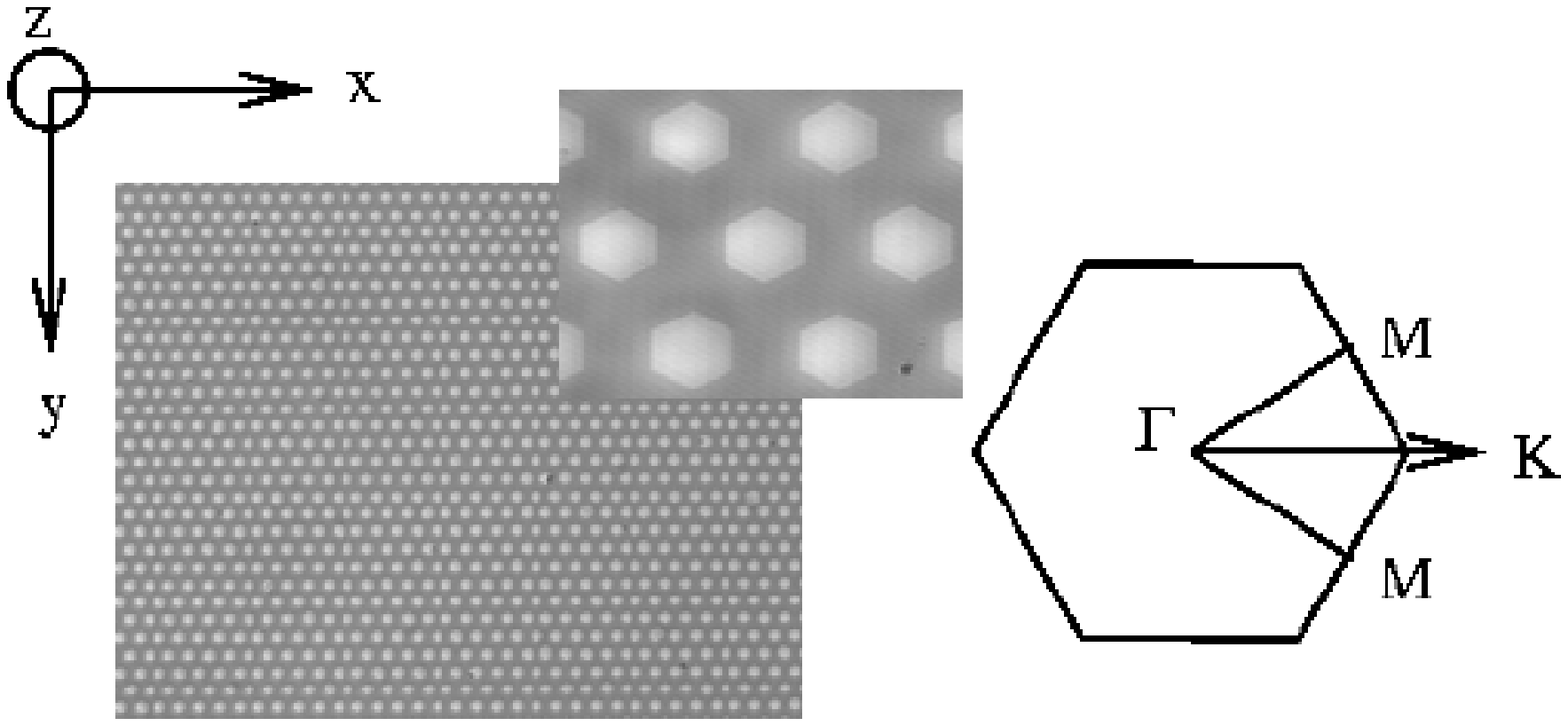}
}
\end{figure}

\newpage

\begin{figure}[h]
\centerline{
\psfig{file=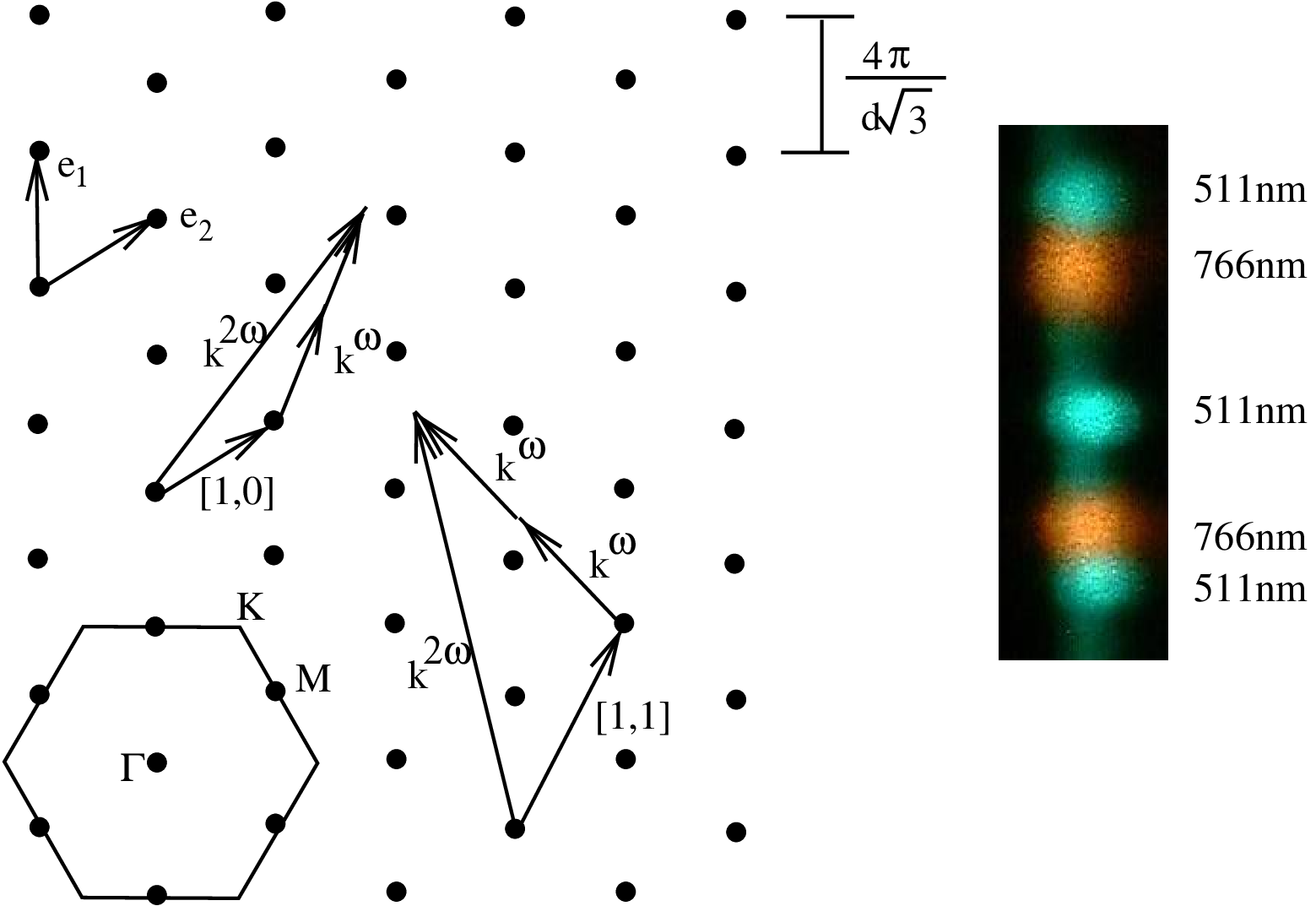,width=8cm}
}
\end{figure}

\newpage

\begin{figure}[h]
\centerline{
\psfig{file=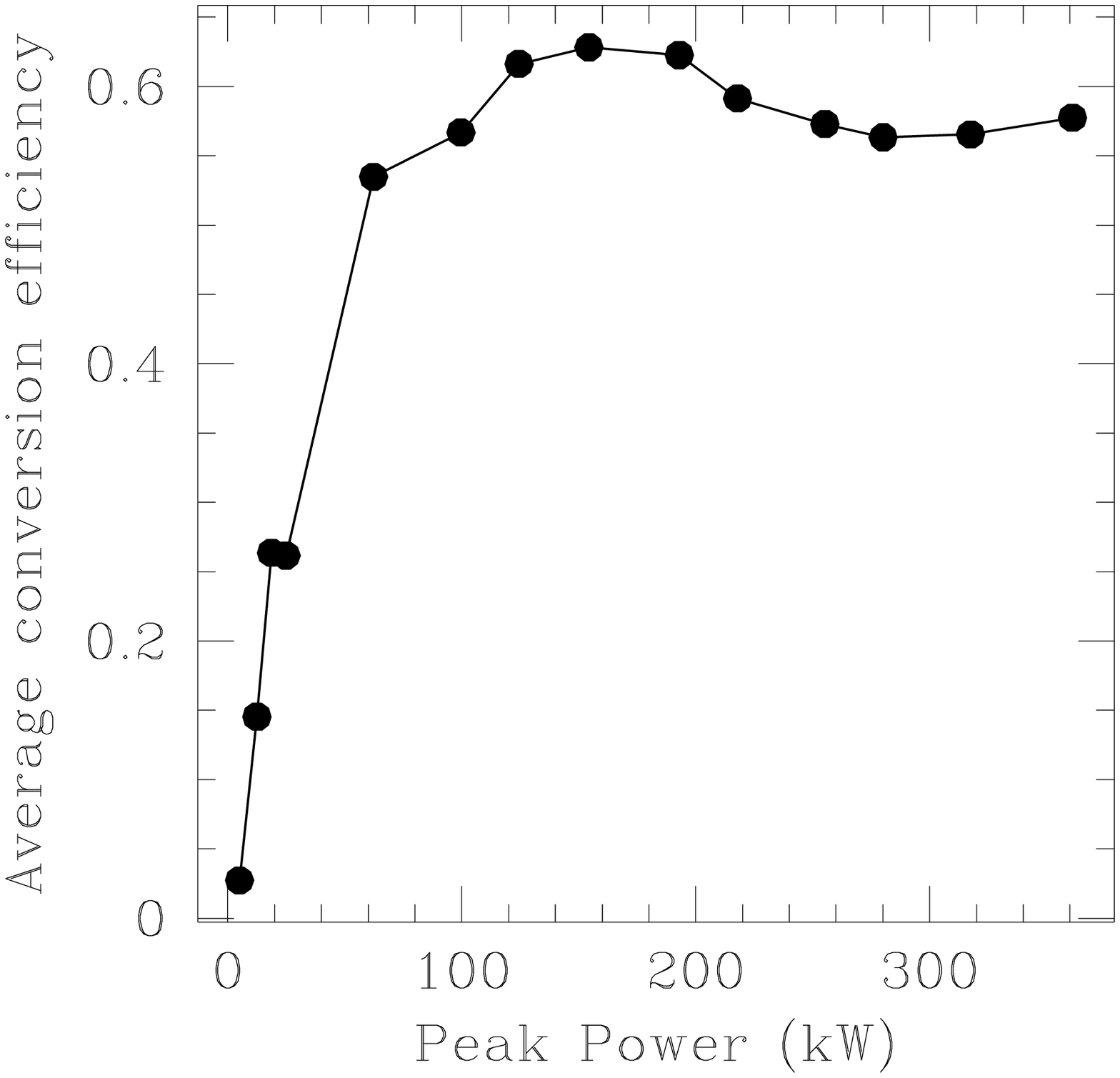}
}
\end{figure}

\newpage

\begin{figure}[h]
\centerline{
\psfig{file=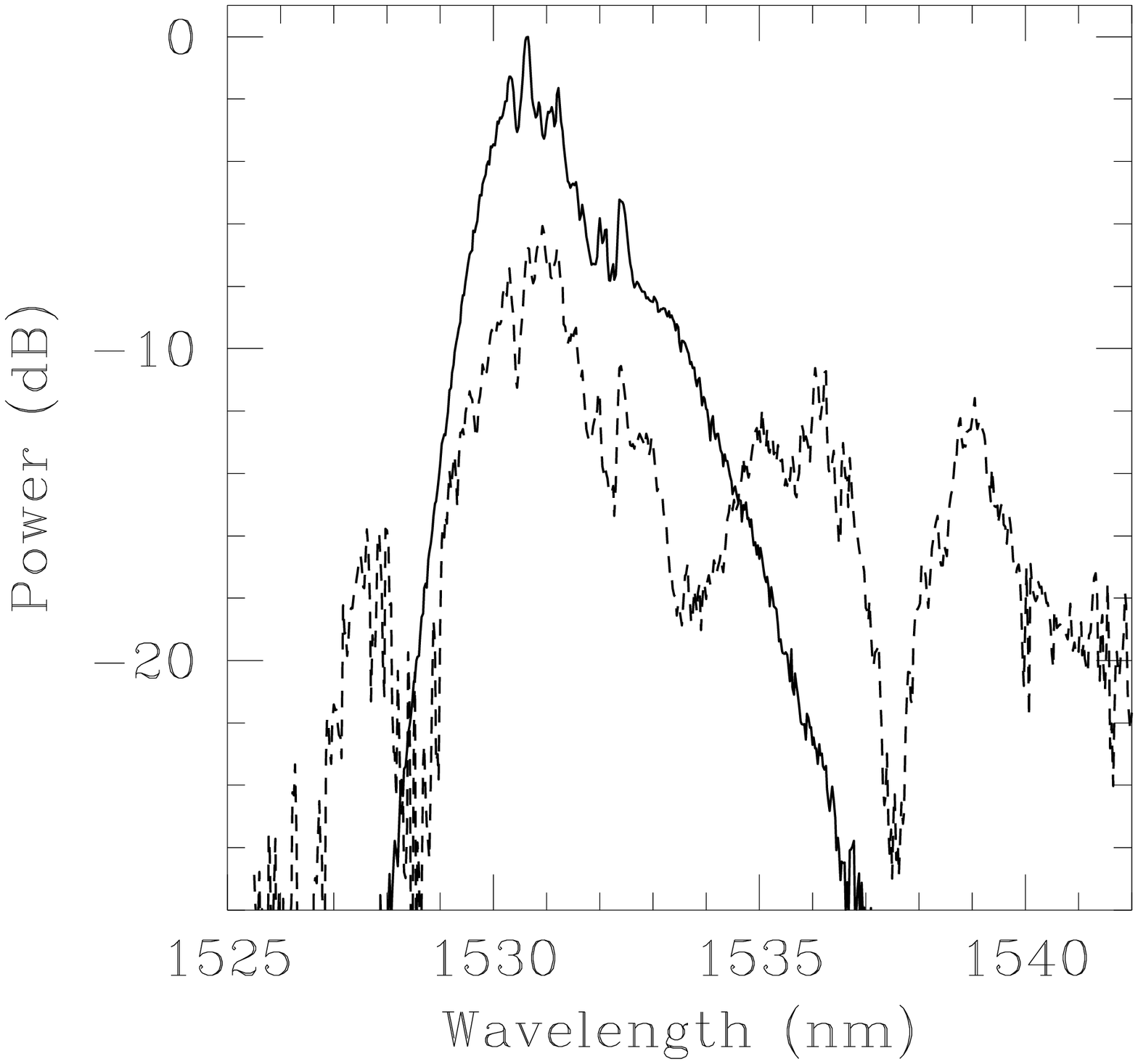}
}
\end{figure}

\newpage

\begin{figure}[h]
\centerline{
\psfig{file=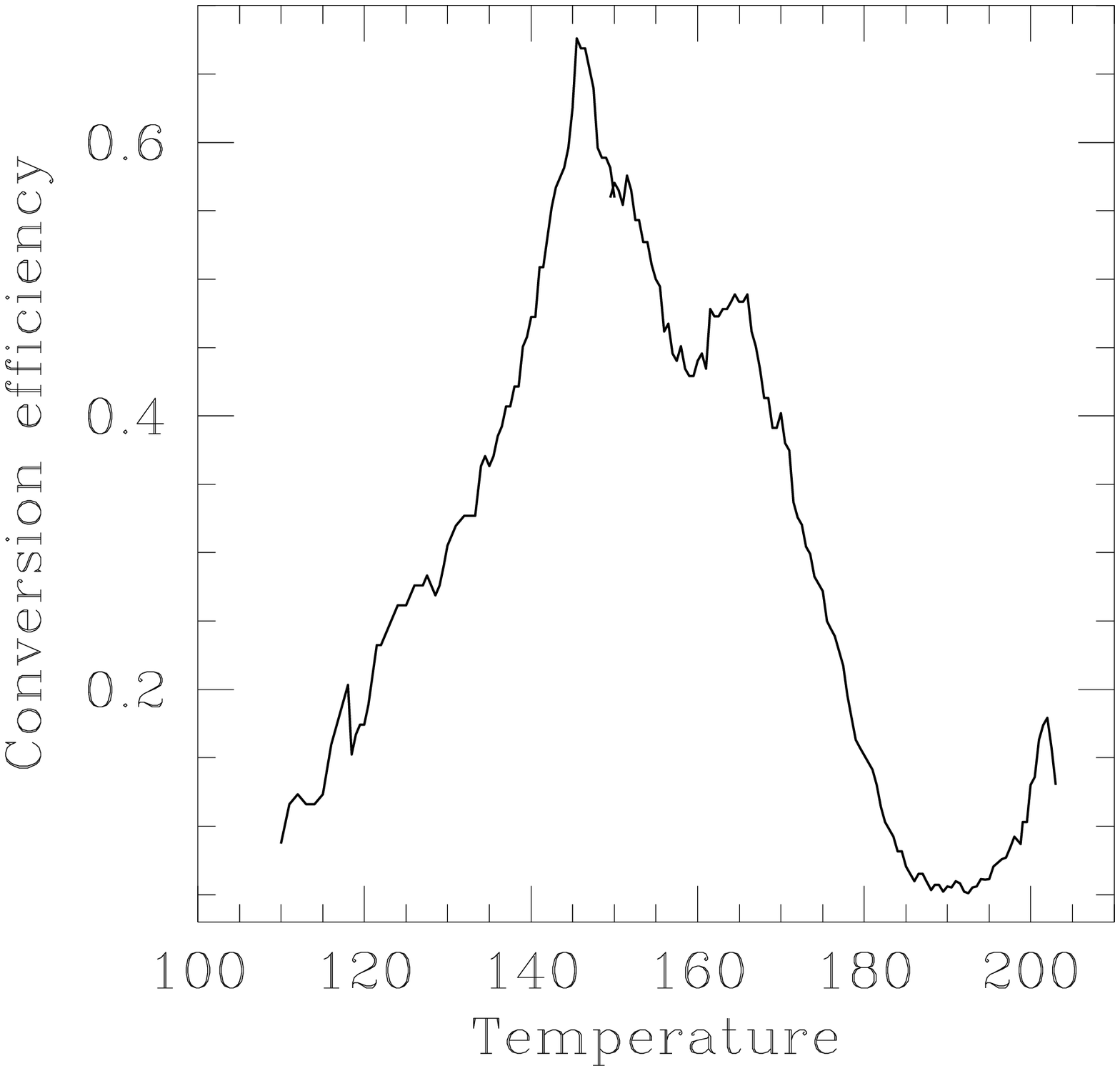}
}
\end{figure}

\newpage

\begin{figure}[h]
\centerline{
\psfig{file=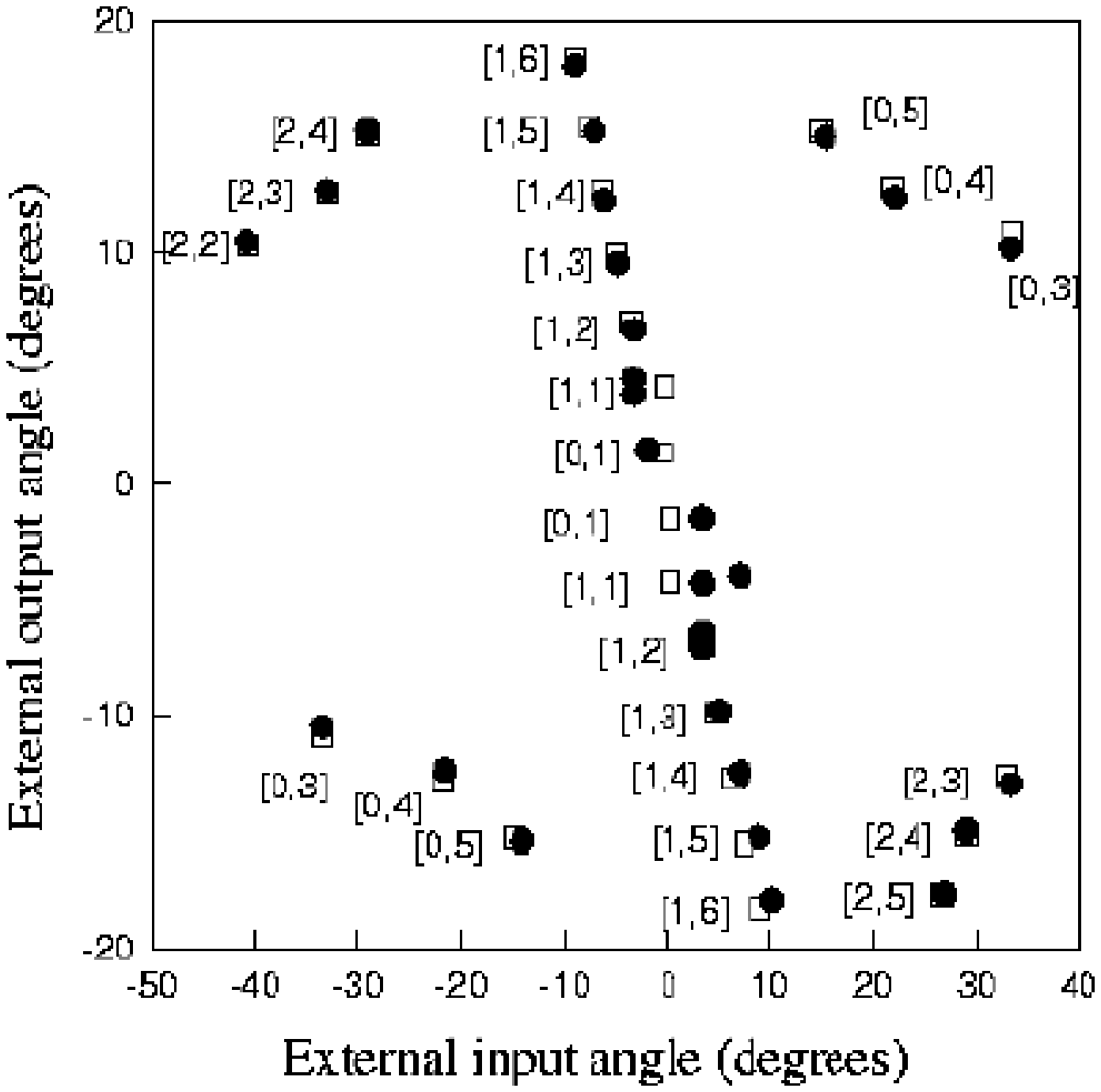}
}
\end{figure}

\newpage

\begin{figure}[h]
\centerline{\psfig{file=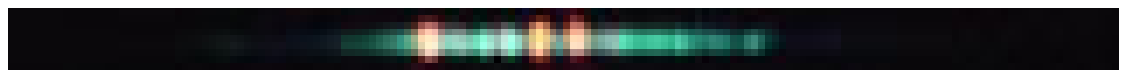,width=8cm}}
\centerline{\psfig{file=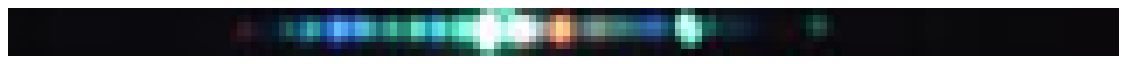,width=8cm}}
\centerline{\psfig{file=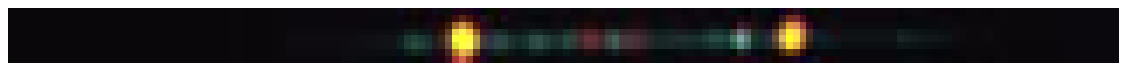,width=8cm}}
\centerline{\psfig{file=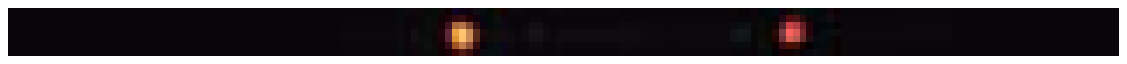,width=8cm}}
\centerline{\psfig{file=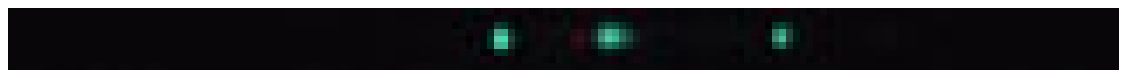,width=8cm}}
\centerline{\psfig{file=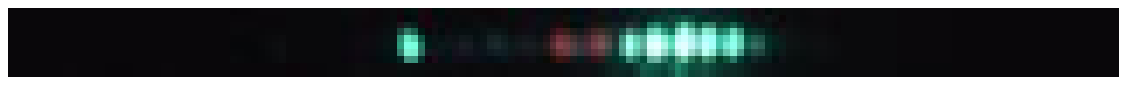,width=8cm}}
\centerline{\psfig{file=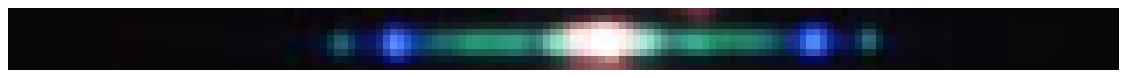,width=8cm}}
\end{figure}

\end{document}